# Beam Distortion Effects upon focusing an ultrashort Petawatt Laser Pulse to greater than $10^{22}$ W/cm$^2$

G. Tiwari,[1,*] E. Gaul,[1,2] M. Martinez,[3] G. Dyer,[3] J. Gordon,[1] M. Spinks,[1] T. Toncian,[4] B. Bowers,[1] X. Jiao,[1] R. Kupfer,[1] L. Lisi,[1] E. McCary,[1] R. Roycroft,[1,5] A. Yandow,[1] G. D. Glenn,[1] M. Donovan,[1] T. Ditmire,[1,2] and B. M. Hegelich[1,6]

[1]Department of Physics, Center for High Energy Density Science, University of Texas at Austin, Austin, TX, 78712, USA
[2]National Energetics, 4616 West Howard Lane, Austin, TX 78728, USA
[3]Standford Linear Accelerator Laboratory, Menlo Park, CA, 94025, USA
[4]Institute of Radiation Physics, Helmholtz Zentrum Dresden-Rossendorf, Dresden, 01328, Germany
[5]Los Alamos National Laboratory, Los Alamos, NM 87545, USA
[6]Center for Relativistic Laser Science, Institute for Basic Science, Gwangju, 61005, South Korea
*Corresponding author: gtiwari30@physics.utexas.edu

**When an ultrashort laser pulse is tightly focused to a size approaching its central wavelength, the properties of the focused spot diverge from the diffraction limited case. Here we report on this change in behavior of a tightly focused Petawatt class laser beam by an F/1 off-axis paraboloid (OAP). Considering the effects of residual aberration, the spatial profile of the near field and pointing error, we estimate the deviation in peak intensities of the focused spot from the ideal case. We verify that the estimated peak intensity values are within an acceptable error range of the measured values. With the added uncertainties in target alignment, we extend the estimation to infer on-target peak intensities of $\geq 10^{22}$ W/cm$^2$ for a target at the focal plane of this F/1 OAP.** © 2018 Optical Society of America

http://dx.doi.org/10.1364/OL.99.099999

Since the advent of chirped-pulse amplification [1] and high power lasers, a significant amount of theoretical and computational research has been carried out to comprehend the fundamental physics of high field (electric field amplitude $\geq 10^{15}$ Volts/m) laser-matter interaction [2]–[4]. The Radiation Reaction (RR) effect, which has eluded experimentalists for over a century since its first description by H. A. Lorentz (1904), is but one example of these strong field interactions [5]. For RR effects to be non-negligible in a laser-matter interaction, the laser intensity must be greater than $10^{22}$ W/cm$^2$ [2, 3]. However, the generation and characterization of laser intensities of $10^{22}$ W/cm$^2$ or greater has been a major challenge for decades. The next generation multi-Petawatt laser projects including the Extreme Light Infrastructure [6], Apollon Laser [7], Laser Project at Shanghai [8], and Aquitaine Laser [9] and existing laser facilities such as the Texas Petawatt Laser facility [10] and the 4 Petawatt laser at CoRELs [11], have the potential to delivering $10^{22}$ W/cm$^2$ on-target. For high power laser systems, it is common to infer the on-target laser parameters from the reference measurements up in the laser chain due to the strong laser-plasma couplings upon target irradiation by intense laser pulses. For example, a peak laser intensity of 7 ×$10^{21}$ W/cm$^2$ was achieved by tightly focusing a 45-TW laser beam with an F/0.6 OAP in 2004 after correction of the residual aberrations with a wave front sensor [12]. Depending on the space-time coordinates of the reference measurement in the beam path, there are limitations to applying a single correction factor to predict peak laser intensities because other factors including, but not limited to, pulse front tilt [13], mid-spatial scattering [14, 15], and pointing and target alignment errors, also contribute to the degradation of the focused beam profile. In this article, we attempt to address the contributions of these effects on the tight focusing of the Texas Petawatt Laser (TPW) by an F/1 OAP.

Based on the combination of optical parametric chirped pulse amplification and mixed neodymium-doped glass, the TPW has a central wavelength of 1.057 μm and was operated with the full width at half maximum (FWHM) pulse duration of 150±20 fs and energy on the range of 80-120 Joules on full power during the F/1 OAP campaign reported here. The F/3 OAP in the target chamber 1 (TC-1) of the TPW facility [10] was replaced by an F/1 OAP (260 mm by 240 mm, segment focal length of 220 mm and off-axis angle of 47º) built by Aperture Optical Sciences to reach on-target intensities beyond $10^{21}$ W/cm$^2$. Prior to the intensity upgrade at TC-1, an additional full aperture (22 cm diameter) adaptive optics system with a built-in iterative feedback loop system was added in the laser chain before the compressor to enable better focusing of the TPW pulse [16].

We report the generation and characterization of peak laser intensities $\geq 10^{22}$ W/cm$^2$ using an F/1 OAP at TC-1 of the Texas Petawatt facility. We identify the effects of residual aberrations, the spatial profile of the near field intensity map, and pointing fluctuations and apply a reduced model to map the corresponding focused spots and their peak intensities from the OSP to TC-1. We also include uncertainties in the target alignment to extend the reduced model and estimate the peak laser intensity encountered by a target at the focal plane of F/1 OAP. We conclude our findings by suggesting methods to improving the quality and repeatability of the focused spot from F/1-like tight focusing.

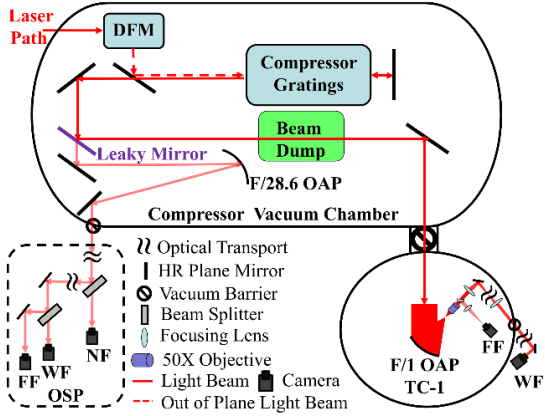

Fig. 1. Schematic of the TPW beam path through the compressor chamber to OSP and TC-1. FF refers to far field, NF refers to near field, WF refers to wave front, DFM refers to the deformable mirror and OSP refers to the Output Sensor Package. Figures are not drawn to scale.

Figure 1 shows the schematic of the TPW beam path from the compressor chamber to the OSP and TC-1. The full aperture deformable mirror (DFM) (250 mm by 220 mm optimized for an angle of incidence of 22.5°), built by ISP System, consists of 52 micro mechanical actuators to correct low spatial frequency (≤ 5 cycles over aperture) aberrations of the laser beam. The wave front correction at the OSP was obtained by running an adaptive optics (AO) based optimization from the OSP. The AO software conducts iterative measurements of the wave front and adjusts the voltages on the DFM to reach a steady state close to the aimed (flat, in our case) phase by running a stochastic optimization algorithm to converge the mean root mean square of the phase aberration map to that of the targeted phase map. The incoming TPW beam, after passing through the DFM and a pair of reflective diffraction gratings in the Compressor Vacuum Chamber, gets transported to a mirror which leaks about 0.1% of the incident light; the transmitted beam passes through plane mirrors and off an F/28.6 OAP to the OSP table where it is down-collimated for detection. The OSP diagnostic comprises both far field (FF) and near field (NF) imaging cameras as well as a PHASICS SID-4 wave front (WF) sensor. Measurements of pulse energy, pulse duration (via second order autocorrelation deconvolution assuming a Gaussian temporal pulse profile) and mid field, not shown in Fig. 1, are also performed at the OSP station. In TC-1, a Plan Apo Infinity Corrected Objective (50X, N. A., 0.55) collects the diverging beam after the focus of the F/1 OAP, which was then passed through a one-inch beam splitter cube for simultaneous FF and WF imaging (see Fig. 1). The reflected beam from the beam splitter passed through an achromatic doublet (f = 200 mm) and the FF image at the focal plane was recorded by a monochrome CMOS camera in vacuum. The transmitted part of the beam was relayed onto another PHASICS SID-4 WF sensor via mirrors and two pairs of achromatic doublets outside the target chamber for wave front characterization. We implemented the standard double pass alignment method using a fiber coupled laser source and R-cube to minimize the aberrations due to the misalignments of the imaging system at TC-1 after the F/1 OAP.

Before each full power laser irradiation, an aberration-corrected Optical Parametric Amplified (OPA) beam was obtained by running the AO loop at the OSP with a flat wave front as a reference using the 2.5 Hz frontend pulse. Figures 2(a) and 2(b) show the far field intensity map and aberration phase map measured at OSP respectively after a typical correction. Then, the longitudinal position of the F/1 OAP along the laser propagation direction was optimized until the best focus image was obtained in the FF camera of TC-1; the focal spot thus obtained was also relayed to PHASICS SID-4 WF sensor outside TC-1. Figures 2(c) and 2(d) show the respective measured FF image and the phase aberration map at TC-1 corresponding to the aberration-corrected OPA beam at the OSP. During the preparation of each full power laser shot, additional compensation for aberrations caused by the amplification of the main amplifiers were also applied; these aberrations are reproducible and can be compensated as constant offsets, once measured and identified. In terms of Zernike polynomials [17], the wave aberration of the applied pre-correction in the latest F/1 OAP run at the OSP is given by $W_{pc} = \frac{\sqrt{3}}{5}(2\rho^2 - 1) - \frac{\sqrt{6}}{10}\rho^2 Cos(2\theta)$ which consists of contributions only from defocus and vertical astigmatism.

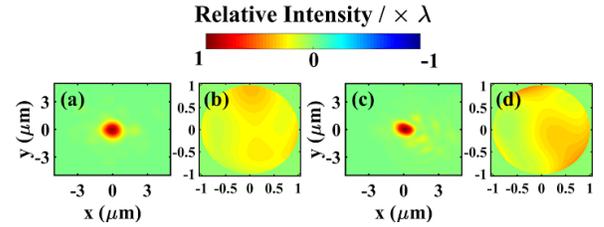

Fig. 2: Measured (a) focused spot by FF camera at OSP, (b) phase aberration map, with mean root mean square (rms) = 0.081λ & Peak-to-Valley (PtV) = 0.36 λ, by SID-4 WF sensor at OSP, (c) focused spot by FF camera at TC-1 and (d) corresponding phase aberration map, with rms = 0.084λ & PtV= 0.49λ, by WF sensor at TC-1 of the aberration corrected OPA beam at OSP. We note that the pupil size of WF sensors at TC-1 and OSP were different in the figures above. The jet colormap depicts relative intensity values of FF map normalized from 0 to 1 and the aberration coefficient values normalized to the laser wavelength (λ).

Figure 3(b) shows the FF profile of the TPW beam recorded at OSP with an energy of 114.8 ± 3.4 Joules and pulse duration of 135 fs with a calculated peak intensity of (2.2 ± 0.07) ×10$^{22}$ W/cm$^2$. To estimate the peak intensity, we apply a Gaussian fit to the profile and use the formula, $Peak\ Intensity = \frac{Peak\ Power}{Effective\ Area\ (EA)} = \frac{Energy}{Pulse\ Duration \times EA}$ where the effective area equals to the area of an circle whose radius encloses 50% of the total energy. The FWHMs are found to be 1.45 µm and 1.2 µm along the x and y axes respectively with an effective radius ($= \sqrt{(FWHMx * FWHMy)}$) of 1.32 µm, which encloses about 56% of the EE as indicated by the brown line in Fig. 4c. Based on scalar diffraction theory, an ideal top-hat TPW beam with similar energy and duration parameters, when fitted to a Gaussian profile, should have a peak intensity of (6.4 ± 0.2)

× $10^{22}$ W/cm$^2$ with an Airy disc intensity profile having the first minimum at a radius of ~1.3 μm (a corresponding FWHM of ~1.05 μm) and the focal tolerance for an F/1 OAP (i.e. intensity drop of 20%) corresponding to ±2.114 μm [17]. The reduction in the measured peak intensity by about a factor of four from the theoretical value indicates the presence of nonideal factors such as residual aberrations, mid-spatial scattering, pulse front tilt, and pointing jitters. We demonstrate some of these effects in terms of deterioration in peak intensity and encircled energies of the FF map by using the measurements correlated to the best measurement of figures 3(a) and 3(b) as an example.

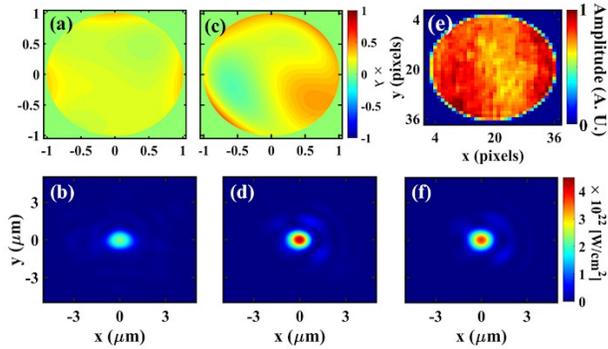

Fig. 3: (a) Aberration phase map measured by WF sensor at the OSP, (b) corresponding FF image recorded at OSP with a record peak intensity of ~2.2×$10^{22}$ W/cm$^2$, (c) Resultant aberration phase map calculated at TC-1 after F/1 focusing and (d) Calculated far field intensity profile based on fast Fourier transform of 3(a) at the focus of F1 OAP. (e) NF amplitude map after background subtraction and pupil adjustment of measured NF intensity map. (f) Calculated focused intensity profile with NF and WF included at the focus of F/1 OAP.

First, we consider the effect of aberrations measured directly by the WF sensor. The presence of residual aberrations is evident based on the difference in aberration phase maps between TC-1 and the OSP (Figure 2). Following the method of Lundström and Unsbo [18] and using the residual aberration map technique [12], we applied a transformation comprising of scaling with a factor of 1.18 and rotation by 140 degrees to the OSP aberration map of Fig. 2(b) to account for the difference in pupil sizes and rotation of the transverse beam profile before subtraction from that of TC-1 (see Fig. 2(d)) to compute the minimal residual aberration between the OSP and TC-1. This residual aberration map has mean root mean square (rms) ~ 0.002λ and Peak to Valley (PtV) ~0.55λ and is added to the transformed aberration phase map (mean rms ~0.07λ & PtV~0.65λ) corresponding to the phase map of Fig. 3(a) at the OSP to obtain the resultant aberration phase map (mean rms ~0.072λ and PtV~ 0.663λ) as shown in Fig. 3(c). Then, we apply a fast Fourier transform (FFT) to the exponential with the aberration, represented by Zernike Polynomials, as its phase over the pupil. The far field intensity profile is represented by the normalized product of the FFT and its complex conjugate for a flat-top intensity profile over the pupil[17]. For the F/1 focusing, the calculated far field profile from the resultant phase map has a peak intensity of (4.2 ± 0.13) × $10^{22}$ W/cm$^2$ (and Strehl Ratio ~ 0.65) with FWHMs of 1.14 μm and 1.09 μm along the x and y axes respectively as shown in figure 3(d). The effective radius of this profile is 1.11 μm and only contains about 60% of the enclosed energy as indicated by the red line in Fig. 4(c). The drop in peak intensity by about 35% and decline of encircled energy compared to the ideal case demonstrate the pronounced effect of aberrations when propagated to the focus of the F/1 OAP. For high power lasers like TPW, it is often tedious and painstaking to get rid of most aberrations due to their low repetition rate at full power.

Second, the mid-spatial scattering (MSS) due to the surface irregularities in OAPs and misalignment of gratings lies below the correction threshold of the adaptive optics package [14, 15]. MSS effects are also absent in the calculated FF intensity map of figure 3(d) above. In addition to MSS, pulse front tilt (PFT) contributes to the irregular intensity distribution of the NF profile measured by a CCD detector. PFTs are also caused by small misalignment of the gratings due to the presence of temporal and spatial chirp in the pulse [13]. Since we could not isolate any of these effects from each other in a recorded near field (NF) intensity map by a CCD camera., we computed the amplitude of the near field profile (Fig. 3(e)) from the measured NF intensity profile at the OSP and applied it along with the aberration map in the FFT calculation to consider these effects. Since the near field profile was not entirely circular, we considered a circular pupil that encompasses most of the near field and used a rectangular region outside the beam profile to get rid of background noise. Then, we calculated the amplitude of the near field profile within that pupil to generate a smooth circular near field profile as shown in Fig. 3(e); this amplitude map is rescaled to the pupil size of the WF sensor and added in the FFT calculation. The new calculated focused spot profile has a reduced peak intensity of (3.7 ± 0.1) × $10^{22}$ W/cm$^2$ as shown in Fig. 3(f). The effective radius of this profile increased to 1.21 μm compared to the earlier case of figure 3(d) and contains about 63.2 % of the energy as shown by blue line in Fig. 4(c).

Furthermore, pointing stability plays an important role in the case of tight focusing scenarios. For instance, the analysis of pointing errors of full power laser shots from the far field data at the OSP indicates standard deviations of 23.4 microradians (μrads) along the major axis and 16 μrads along the minor axis as indicated by the blue ellipse shown in Fig. 4(a). Simulations conducted on Zemax OpticStudio for an ideal Top Hat TPW beam with a peak power of 0.851 PW and 19.2 μrads pointing error along the y-axis indicate a drop in the peak intensity by about 26% compared to the ideal case. For the case of pointing error with 17.5 μrad along the x-axis and 22.7 μrads along the y-axis, we observe a drop in peak intensity by about 51% (see Fig. 4(b)). Similarly, the encircled energy drops significantly when pointing errors are present as indicated by purple and magenta lines shown in Fig. 4(c). The effect of the latter pointing error seems to be strongest among the recognized distortion factors; within 1 μm radius, the enclosed energy profile of simulated TPW beam with pointing error of (17.5, 22.7) μrad (magenta line) approaches that of the FF measured at OSP (brown) as shown in Fig. 4(c).

In addition, there are often alignment errors in the placement of the targets at the focus of F/1 OAP caused mainly by the dimensional tolerances of the target holder, the roughness of the target surface and the minimum step movement of the translation motor. At the TPW facility, we typically used a Micronix MP-21 with 2.5 μm per step for target alignment; this induces fluctuations of about a 30% drop in the peak intensity per step due to defocus effects [17]. Similarly, the target surface roughness induces negligible defocus and pointing errors.

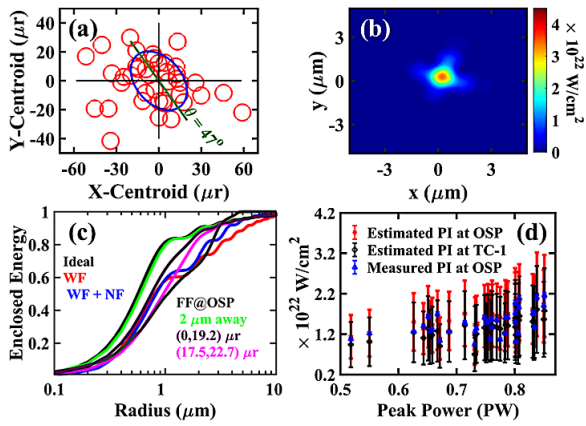

Fig. 4: (a) Pointing error analysis of Full Power FF measured data at OSP in the F/1 OAP campaign; each red solid circular ring represents a focused spot data with the radius equivalent to the 1.3 µm and with the centroid located at the origin. (b) Calculated FF profile at the focus of F/1 OAP with pointing error of (17.5, 22.7) µrads in Zemax Opticstudio. (c) Enclosed Energy profiles for various cases versus the radius (µm) as a distance from the peak intensity spot of each FF map. (d) Plot of Peak Intensities (PI) measured at OSP and calculated using reduced model at OSP and TC-1 versus Peak Power (PW) as shown in Table 1.

Based on the above analysis, the presence of beam distortion factors is marked by an obvious drop in Strehl ratio and peak intensity. Using the best FF measurement at OSP, we can now use a reduced model to predict peak laser intensities at OSP; Table 1 shows the scheme with the list of the included distortion factors and their effects on the quality of focused TPW beam in terms of Strehl ratio and peak intensity. As shown in Fig. 4(d), the estimated values (red) obtained from the reduced model agree with the measured values (blue) at OSP within the permissible error range anticipated by the presence of the beam distortion factors. We estimate on-target intensities of ≥$10^{22}$ W/cm$^2$ from the F/1 OAP at TC-1 based on this scheme as indicated by black data points in Fig. 4(d)).

**Table 1. Reduced Model Scheme for Estimating on-target Intensity at the focus of F/1 OAP at TC-1**

| Beam Factors | Strehl Ratio | Peak Intensity (× $10^{22}$ W/cm$^2$) |
|---|---|---|
| 0.85 PW in F/1 focus |  | 6.4 |
| Strehl upto 8$^{th}$ Zernike | × 0.8 | ~5.1 |
| **Adding Realistic Beam Distortions** | | |
| Scattering/mid spatial | × 0.8 | ~4.1 |
| Pulse Front Tilt | ~× (0.9-1) | ~ (3.89±0.1) |
| Relative Pointing Jitter | ~× (0.5 – 0.8) | ~ (2.5±1.2) |
| **Adding Target Alignment Errors** | | |
| Z-axis error of ± 2.5 µm | × (0.7 – 1) | ~ (2.15±1.3) |
| Tilted Target Error | ~× (0.95 – 1) | ~ (2.1±1.23) |

Overall, we have shown that although the TPW can generate intensities of ≥$10^{22}$ W/cm$^2$ for a target at the focal plane of the F/1 OAP, the quality of the focused spots deviates significantly from the ideal case when the effects of realistic beam distortion factors and target alignment errors are considered. Some modifications can help improve the quality of tightly focused spots. For instance, mid-spatial scattering can be avoided by using highly polished OAP mirrors in the laser chain [19] and pulse front tilt (PFT) can be adjusted via alignment of gratings in the laser chain. PFT can be measured, but doing so will be difficult on the TPW since the measurement method is scanning-based and requires high repetition rate at high power [20]. Improved optic mounts and active, close-loop stabilization systems with high repetitive pilot beams can improve pointing stability. High precision motor stages and the Confocal High Intensity Positioner method [21] should be adopted to avoid alignment errors on the target side after F/1 focusing.

**Funding.** Air Force Office of Scientific Research (FA9550-14-1-0045, FA9550-17-1-0264); Defense Advanced Research Project Agency Contract 12-63-PULSEFP014; National Nuclear Security Administration Cooperative Agreement DE-NA0002008.